\documentclass[prb,twocolumn,showpacs,floatfix]{revtex4}
\usepackage{graphicx}
\usepackage{dcolumn}
\usepackage{bm}
\usepackage{epstopdf}
\newcommand{\p}{\\[2ex]}

\newcommand{\sstrut}{\rule[-1ex]{0ex}{3.5ex}}
\newlength{\figwidth}
\newlength{\hw}
\newlength{\vvp}
\newlength{\minusspace}
\newcommand{\msp}{\hspace{\minusspace}}
\newlength{\zerospace}
\newcommand{\degree}{$^\circ$}
\newcommand{\mub}{$\mu_{\mathrm B}$}

\newcommand{\Tc}{T$_{\mathrm{C}}$}
\newcommand{\Tcn}[1]{T$_{\mathrm{C}#1}$}
\newcommand{\Tn}{T$_{\mathrm{N}}$}
\newcommand{\hf}{\ensuremath{\frac12}}
\newcommand{\gdc}{Gd$_2$CuO$_4$}
\newcommand{\Sv}[1]{\ensuremath{{\bf S}_#1}}
\newcommand{\Tv}[1]{\ensuremath{{\bf T}_#1}}
\newcommand{\bv}{\ensuremath{\bf b}}
\newcommand{\av}{\ensuremath{\bf a}}
\newcommand{\cv}{\ensuremath{\bf c}}
\newcommand{\subs}[1]{\ensuremath{_{\mbox{\scriptsize #1}}}}
\newcommand{\pv}{\ensuremath{{\bm \tau}}}
\begin{document}
\title{Weak ferromagnetism and magnetic phase transitions in Gd$_2$CuO$_4$}
\author{P.J. Brown and T. Chatterji }
\address{Institut Laue-Langevin, B.P. 156, 38042 Grenoble Cedex 9, France\\ }
\date{\today}
\begin{abstract}
We report a polarised neutron study of the magnetic structures and phase trasitions in \gdc\ in low magnetic fields. These experiments 
have been complemented by integrated intensity measurements with unpolarised neutrons in zero field.  Polarised neutron flipping ratio measurements have been made with magnetic fields $H = 0.05$, $0.10$ and 0.5~T in the temperature range 4-20~K. These have enabled us to deduce that the anomalous temperature behaviour of the coherent magnetic scattering from the Cu sublattice, which shows sharp  intensity minima at \Tcn1 $\approx 18$K and \Tcn2 $\approx 8$ K, is due to cross-overs in the sign of the interaction between strongly coupled, weakly ferromagnetic, CuO$_2$ layers. At \Tcn1\ the coupling changes from ferromagnetic to anti-ferromagnetic and long-range order between layers is temporarily lost. \Tcn2\  is the temperature at which the Gd moments order and a further reorganisation of the interlayer order takes place. The weak ferromagnetism of the CuO layers is found
to be due to a small rotation of the Cu moments in the same direction as that in which their coordinating oxygen squares rotate in the tetragonal to orthorhombic distortion of the crystal structure. Further analysis of the flipping ratio measurements has enabled us to model the magnetic structures of the zero-field and the field-induced phases of Gd$_2$CuO$_4$.
\end{abstract}
\maketitle
\section{Introduction}
The magnetic properties of antiferromagnets that contain two magnetic species  are often very complex due to the presence of several competing exchange interactions. The series  R$_2$CuO$_4$ (R = rare earth element), the parent compounds of the electron-doped superconductors, are examples of such interesting magnetic systems. Studies of  Nd$_2$CuO$_4$ and Pr$_2$CuO$_4$ \cite{matsuda90,skanthakumar93,chattopadhyay91,sumarlin95} have shown that the Cu$^{2+}$ moments order close to room temperature whereas magnetic order in the R$^{3+}$ sublattice is only established at very low temperatures. However several spin-reorientation transitions which have been observed at intermediate temperatures, notably in Nd$_2$CuO$_4$, demonstrate competition between Cu-Cu, R-Cu and R-R exchange.  Gd$_2$CuO$_4$ is unique amongst these  compounds in that although it is as easily doped as other members of the series, it does not, as they do, become superconducting. \cite{markert:89}  Additionally, it shows weak ferromagnetism below the Cu ordering temperature $T_N$ (Cu) $\approx 285$~K \cite{thomson:89,seaman:89,xiao:89,oseroff:89}.  
The large absorption cross-section for thermal neutrons of natural Gd has limited neutron diffraction studies of \gdc. However such studies have been made possible by the availability of $^{158}$Gd enriched Gd$_2$CuO$_4$ single crystals and the magnetic structures of the Cu and Gd sub-lattices have been determined \cite{chatt:91,chatt:92}.
 \subsection{Crystal and Magnetic Structure}
\begin{figure}[htbp]
\begin{center}
\resizebox{0.4\textwidth}{!}{\includegraphics{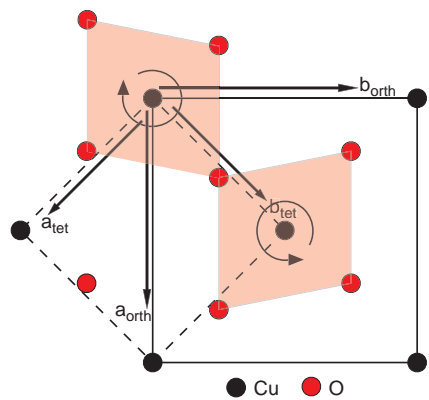}}
\caption{The layer at z=0 of the orthorhombic structure of \gdc\ showing the rotation and distortion of the oxygen squares which coordinate the Cu$^{2+}$ ions. }
\label{orthcell}
\end{center}
\end{figure}
The crystal structure of \gdc\ at ambient temperature is derived from the tetragonal Nd$_2$CuO$_4$ structure by a small orthorhombic distortion. \cite{braden:94}
At \Tn\ = 285 K
the Cu$^{2+}$ moments order in the  Nd$_2$CuO$_4$
type of antiferromagnetic structure with magnetic moments parallel to the propagation vector $\pv=(\frac{1}{2}, \frac{1}{2},0)\subs{tet}$ \cite{chatt:92}. The Gd moments order at a much lower temperature  $T_N = 6.4$ K\cite{chatt:91} to a structure with zero propagation vector in which  oppositely oriented moments lie in the \av-\bv\ plane and are related by the centre of symmetry. 

The weak ferromagnetism observed below \Tn\ (Cu) in \gdc\ is not compatible with a magnetic structure with non-zero propagation vector (vis \hf\hf 0\subs{tet}) but is allowed in the orthorhombic cell for which the propagation vector \hf\hf0\subs{tet}=100\subs{orth} is a lattice vector. The conventional space group of the orthorhombic structure is
Cmca, however in order to retain the c-axis of the tetragonal cell it is convenient to describe it  using the space groups  Acam and Bbcm for the two twins that can be formed from the tetragonal phase.
$\av\subs{orth}=\av\subs{tet}\pm\ \bv\subs{tet};\ \bv\subs{orth}=\av\subs{tet} \mp\ \bv\subs{tet}$ 
 $\cv\subs{orth}=\cv\subs{tet}$.
The main effect of the structural distortion is to rotate the squares of O$^{2-}$ ions so that, as shown in figure~\ref{orthcell}, these ions no longer lie on the $[110]$ and $[1\bar10]$ axes of the orthorhombic cell.  In the A face centred cell (A twin) the Cu layer at $z=\hf$ is displaced  by \hf\bv\ 
with respect to that at $z=0$ and in the B face centred cell (B twin)  by 
\hf\av. 
 Magnetic domains with propagation vector (\hf\hf0)\subs{tet}  belong to the A twin  and those  with propagation vector (\hf-\hf0)\subs{tet}  to the B twin. The O$^{2-}$ squares coordinating Cu$^{2+}$ ions with antiparallel moments are rotated in opposite directions 
 allowing weak ferromagnetism.
 
The orthorhombic cell will be used for axes and reflection indices throughout the rest of this paper.

The Bragg reflections from the orthorhombic structure of \gdc\ are classified in table~\ref{sfacs} according to the 
nuclear contribution to their structure factors. 
\begin{table}
\setlength{\hw}{-1.5ex}
\setlength{\vvp}{0.8ex}
\caption{Conditions on the reflection indices $h, k$ and $l$ for nuclear scattering by the orthorhombic structure of \gdc.}
\begin{tabular}{llcll}
\hline
&$h+l$&\\[\hw]
Indices&&Type&\multicolumn{1}{c}{Description}\\[\hw]
&$k+l$\\
\hline
&even&&Fundamental: given by\\[\hw]
$hkl$&&F\\[\hw]
&even&& the tetragonal structure\\[\vvp]
&even&&Given by B-face centred \\[\hw]
$hkl$&&B\\[\hw]
&odd&&twin only\\[\vvp]
&odd&&Given by A-face centred\\[\hw]
$hkl$&&A\\[\hw]
&even&&twin only\\[\vvp]
&even\\[\hw]
$h0l$&&SB&Space group absence\\[\hw]
&odd\\[\vvp]
&odd\\[\hw]
$h0l$&&SA&Space group absence\\[\hw]
&even\\[\vvp]
\hline
\label{sfacs}
\end{tabular}\\[1ex]
\end{table}
 \subsection {Magnetic phase diagram}
 \begin{figure}[htbp]
	\centering{
 	 \resizebox{0.45\textwidth}{!} {\includegraphics{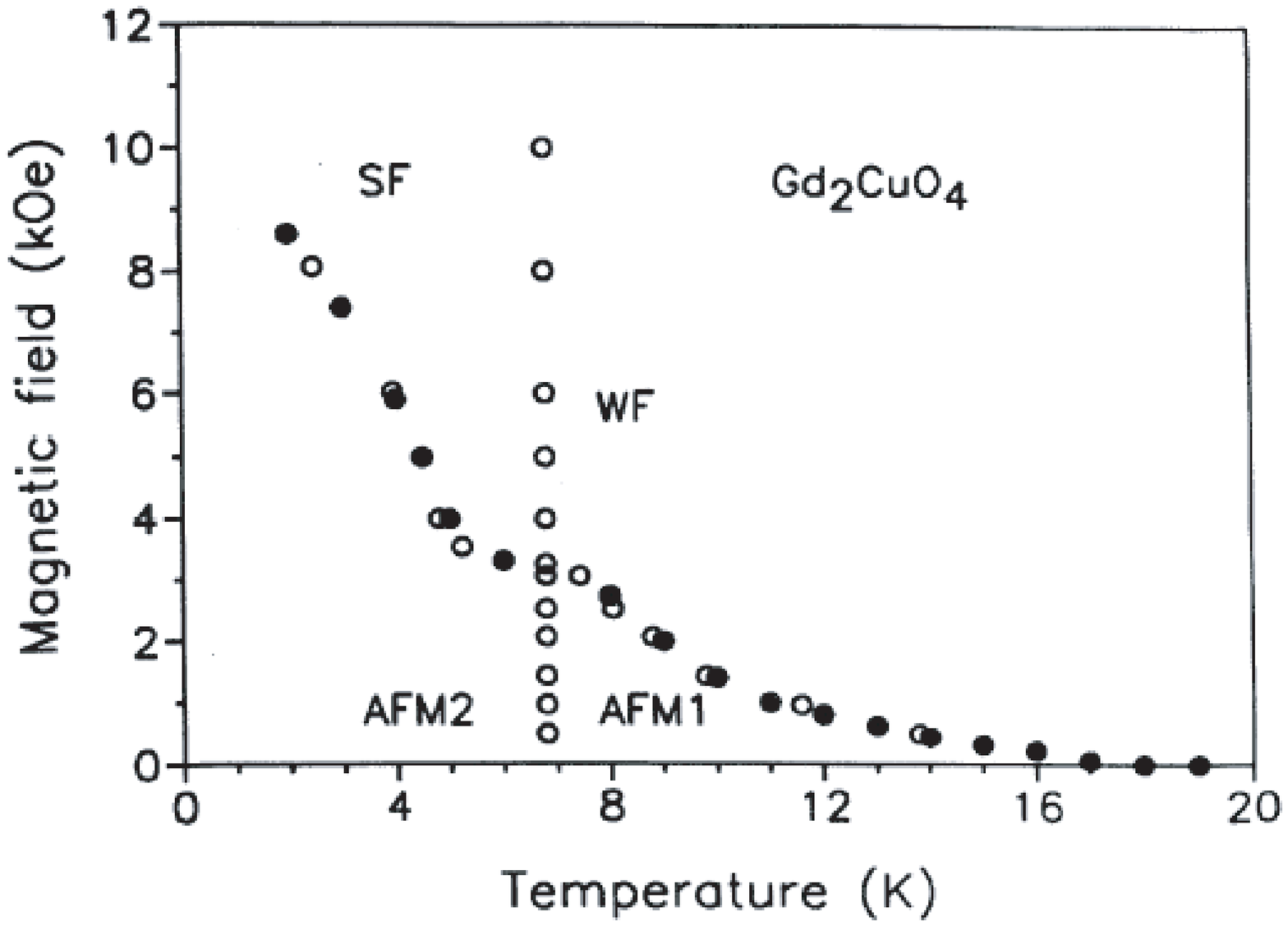}}   
   \caption{H vs T Phase diagram of \gdc\ \cite{stepanov:93}
   \label{htphases}}
  }
\end{figure}
A tentative magnetic phase diagram for \gdc\ based on magnetisation and neutron diffraction results \cite{stepanov:93} is shown in 
figure~\ref{htphases}. The weak ferromagnetic phase (WF) is stable from \Tn\ (Cu) down to $\approx 20$~K in zero field, at which temperature the weak ferromagnetic moment drops to zero (AFM1 phase). The WF phase can be retained down to 7~K by application of rather small magnetic fields $\approx 0.2$~T at 10~K. Below 6.4~K both the Cu and The Gd moments are ordered giving the AFM2 phase which undergoes a spin flop transition to the SF phase on raising the field.
The variation with temperature of the integrated intensities of the 101 and 102 reflections to which the ordered copper moments contribute strongly  show two anomalous minima which coincide approximately with the WF to AFM1 and AFM1 to AFM2 transitions.\cite{stepanov:93} The object of the present experiment was to study the low field behaviour of the magnetic structure using  polarised neutron diffraction
in order to clarify the complex interplay between the 
order parameters of the rare-earth and the transition metal sub-lattices which  leads to the appearance and subsequent disappearance of weak ferromagnetism and to the apparent breakdown of order in the copper sub-lattice as the Gd sub-lattice orders. 

\section{Experimental}
The polarised neutron measurements were made using the diffractometer D3 which uses a spin polarised neutron beam from the hot-source of the high flux reactor of the ILL Grenoble.
The \gdc\ crystal, of  size $\approx 5\times8\times2$~mm$^3$ was a piece of the one used for the magnetic structure determination\cite{chatt:91,chatt:92}.  It was mounted in a thin tailed cryostat which can pass
 through the hollowed out soft iron poles of the D3 electromagnet. An [010] axis was aligned parallel to the field direction. Initially the crystal was cooled in zero field and the integrated intensities of the 101 and 102 reflections measured as a function of temperature between 23 and 2~K. A magnetic field of  0.05~T was then applied which is sufficient to ensure 82\%
 neutron polarisation. The flipping ratios $R$ of two F type reflections  \{111\} and \{113\},  one SA type \{101\},  one SB type \{102\} and two each of the A and B types  \{210\}, \{212\} and \{211\}, \{213\} were measured in the same temperature range. Most of the measurements were repeated in fields of 0.1 and 0.5~T.
The integrated intensities of the SA and SB reflections 101 and 102 were also measured before and after the 
polarised neutron experiment using the D9 4-circle diffractometer which is also on the ILL hot-source.  

\section{Results}
\begin{figure*}[htbp]
    \centering{
  	 \resizebox{\textwidth}{!} {\includegraphics{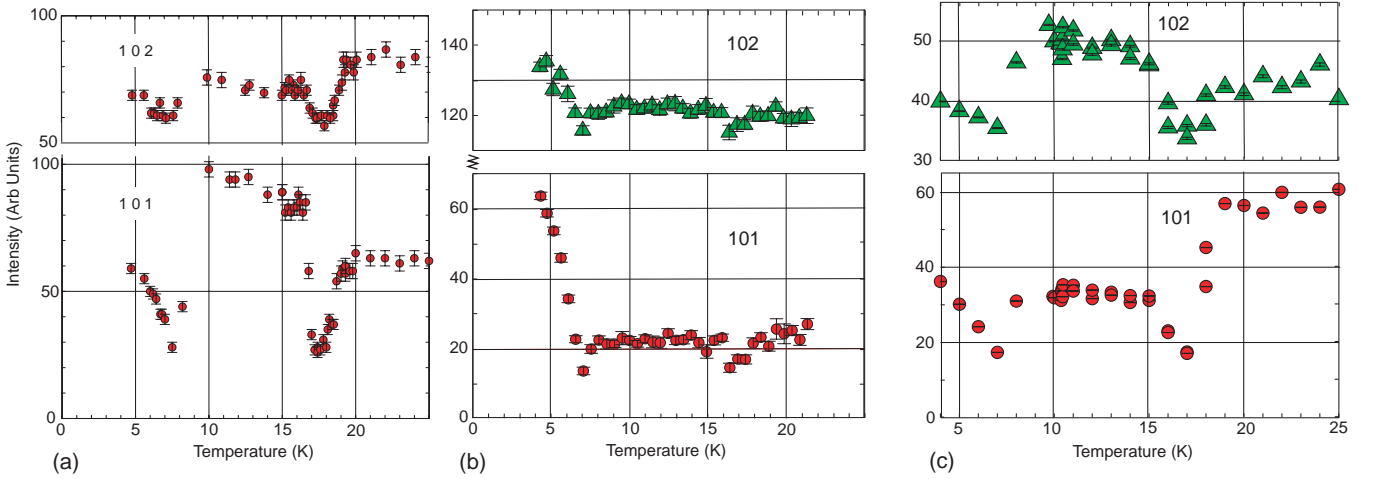}}
}	 
    \caption{Temperature variation of the integrated intensities of the 101 
   and 102 magnetic reflections from \gdc\ from 2 to 22 K (a) Initial experiment on D9, (b) Measurements made on D3, (c) Second experiment on D9}
    \label{intvst}
 \end{figure*}

\begin{figure*}[htbp]
   \centering
 	 \resizebox{\textwidth}{!} {\includegraphics{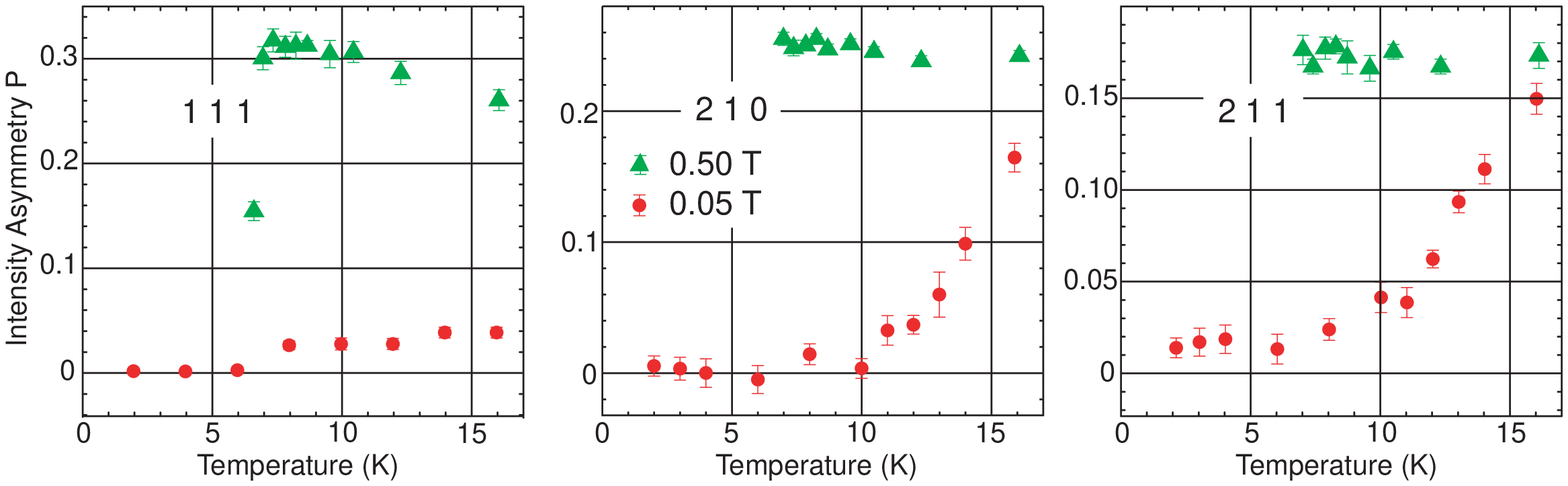}} 
   \caption{Intensity asymmetry $(R-1)/(R+1)$ in the F:111, B:210 and A:211 type
   reflections in 0.05 and 0.5 T $\|$ [010].}
   \label{polvst}
\end{figure*}

The temperature variation of the integrated  intensities measured for the 101 and 102 reflections in the three different experiments is illustrated in figure~\ref{intvst}  In all three the intensity was found to fall abruptly in a small temperature range around \Tc$_1$ and  \Tc$_2$.  Although the breaks in the curves occur at the same two temperatures  the relative intensities at other temperatures are very different suggesting that the magnetic intensity in a particular reflection depends on the previous history of the sample. 
It was found that that the anomalous breaks in intensity can be suppressed by magnetic fields of as little as 1~T. 
\p 
 A polarisation dependence of the scattered intensity is expected due to the weak ferromagnetism associated with the structural modulation which
accompanies the N\'eel transition. This modulation has the same wave-vector as the magnetic structure so that the {\em magnetic} reflections
are not entirely of magnetic origin \cite{ozer:90}.  In the orthorhombic cell the Cu moments are not constrained to lie  parallel to the crystallographic axes and the weak ferromagnetism arises from a small rotation of the moments away from the axes. 
Significant polarisation dependence  was found in the intensities of all reflections except those of types SA and SB for which the nuclear structure factors are zero. The variation of the intensity asymmetry $P = (R-1)/(R+1)$ with temperature, in fields of 0.1 and 0.5~T, is shown in figure~\ref{polvst} for a reflection of each of the F, A and B types.
 At the lowest field
(0.05~T) the asymmetry of the A and B reflections starts to increase rapidly above about 12~K. According to the phase diagram of reference \cite{stepanov:93} this is the phase boundary between the AFM1 and WF phases in 0.05~T, suggesting that the increasing asymmetry is due the presence of the WF phase. The F type reflections 111 and 113 also have significant asymmetry at temperatures above \Tcn2. The rapid loss of asymmetry at lower temperature 
results from the rapid increase in the polarisation independent intensity due to  antiferromagnetic ordering of the Gd sub-lattices.  
\p
\section{Analysis}
\begin{figure}[htb]
   \centering
 	 \resizebox{0.48\textwidth}{!} {\includegraphics{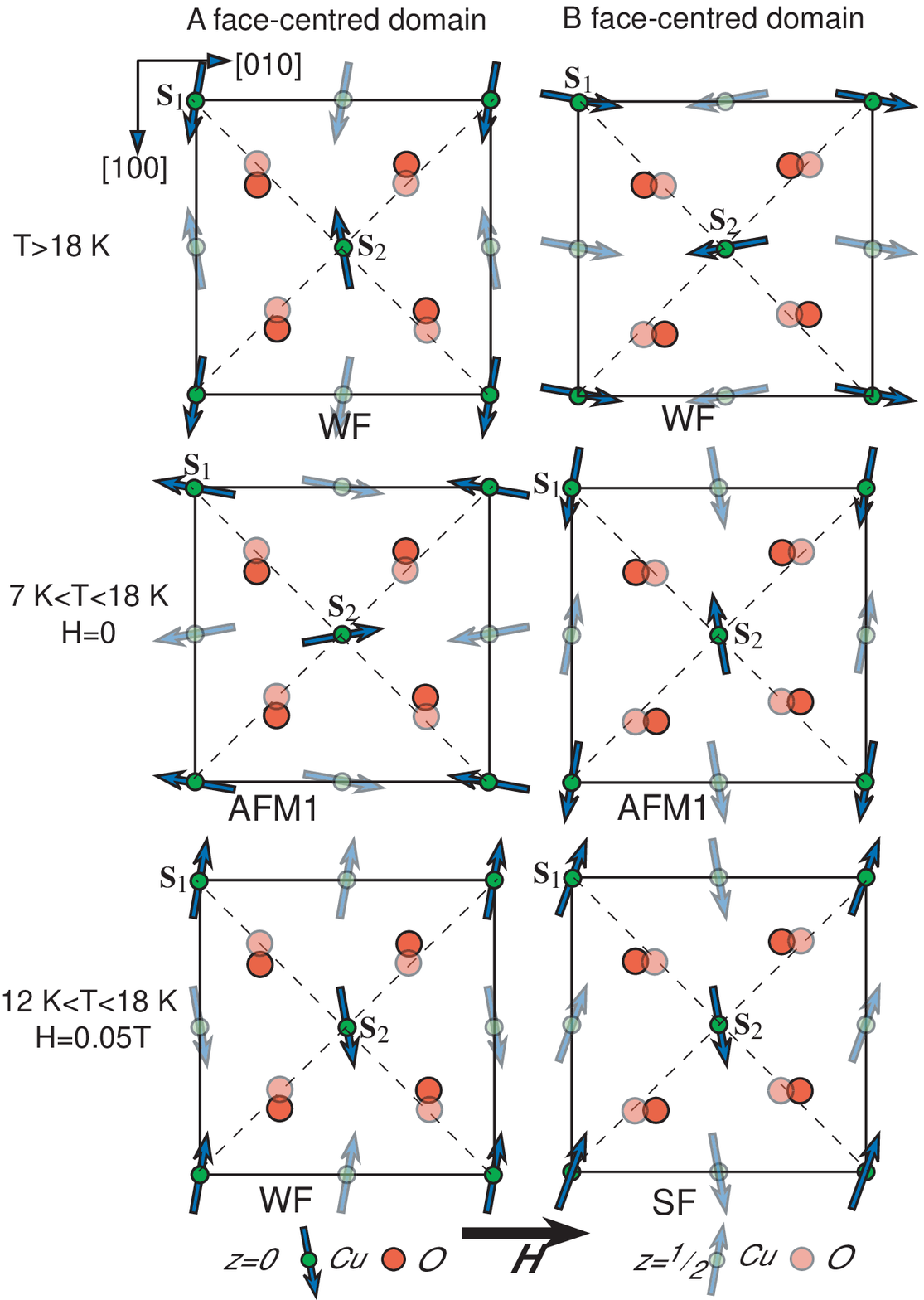}} 
   \caption{Magnetic structures proposed for the WF, AFM1 and SF1 phases of
   \gdc.}
   \label{phases}
\end{figure}

In a centrosymmetric structure like \gdc\ the cross-section can only be polarisation dependent when nuclear and magnetic scattering occur at the same wave-vector and from the same coherent volume of crystal. In the case of the A reflections this magnetic scattering must therefore also come from the A twin, and for the B reflections from the B twin although magnetic scattering from both twins will contribute to the total intensity in the F reflections. For the Cu sub-lattice the magnetic propagation vector on orthorhombic axes is either 100 or 010. Furthermore within the orthorhombic cell the Cu ions at (000)  and (\hf\hf0 )are related by an a glide-plane $\perp$ {\bf b} for the A twin and a 
 b glide-plane $\perp$ {\bf a} for the B twin.
If the magnetic moments on the Cu ions at (000) and (\hf\hf 0) are given by the vectors \Sv1\ and \Sv2\ and those on the Gd ions  at ($0,0,\pm z_{Gd}$) and ($\hf,\hf,\pm z_{Gd}$) by \Tv1, \Tv2, \Tv3 and \Tv4 respectively then the magnetic structure factors of the different types of nuclear reflection are 
\begin{table*}[htb]
{
\caption{Combinations of the magnetic moment vectors of Cu and Gd ions contributing to magnetic scattering in different types of reflections in \gdc.\label{combs}
} 
\setlength{\hw}{-1.5ex}
\setlength{\vvp}{0.8ex}
\begin{center}\begin{tabular}{ccccrccr}
\hline
&&\quad&\multicolumn{2}{c}{A face-centred twin}&\quad&\multicolumn{2}{c}{B face-centred twin}\\[\hw]
Type&$\pv$\\[\hw]
&&&\multicolumn{1}{c}{Cu}&\multicolumn{1}{c}{Gd}&&\multicolumn{1}{c}{Cu}&\multicolumn{1}{c}{Gd}\\
\hline
&&&&$C(\Tv1+\Tv2+\Tv3+\Tv4)$&&&$C(\Tv1+\Tv2+\Tv3+\Tv4)$\\[\hw]
&100&&$\Sv1+\Sv2$&&&$\Sv1+\Sv2$&\\[\hw]
&&&&$+ \imath S(\Tv1-\Tv2+\Tv3-\Tv4)$&&&$+ \imath S(\Tv1-\Tv2+\Tv3-\Tv4)$\\[\hw]
F\\[0.5\hw]
&&&&$C(\Tv1+\Tv2+\Tv3+\Tv4)$&&&$C(\Tv1+\Tv2+\Tv3+\Tv4)$\\[\hw]
&010&&$\Sv1+\Sv2$&&&$\Sv1+\Sv2$\\[\hw]
&&&&$+ \imath S(\Tv1-\Tv2+\Tv3-\Tv4)$ &&&$+ \imath S(\Tv1-\Tv2+\Tv3-\Tv4)$\\[0.5ex]
\hline%
&&&&$C(\Tv1+\Tv2-\Tv3-\Tv4)$\\[\hw]
&100&&$\Sv1-\Sv2$&&&0&\multicolumn{1}{c}{0}\\[\hw]
&&&&$+ \imath S(\Tv1-\Tv2+\Tv3-\Tv4)$\\[\hw]
A\\[0.5\hw]
&&&&&&&$C(\Tv1+\Tv2-\Tv3-\Tv4)$\\[\hw]
&010&&0&\multicolumn{1}{c}{0}&&$\Sv1-\Sv2$\\[\hw]
&&&&&&&$+ \imath S(\Tv1-\Tv2-\Tv3+\Tv4)$\\[0.5ex]
\hline%
&&&&$C(\Tv1+\Tv2-\Tv3-\Tv4)$\\[\hw]
&100&&$\Sv1-\Sv2$&&&0&\multicolumn{1}{c}{0}\\[\hw]
&&&&$ + \imath S(\Tv1-\Tv2-\Tv3+\Tv4)$\\[\hw]
B\\[0.5\hw]
&&&&&&&$C(\Tv1+\Tv2-\Tv3-\Tv4)$\\[\hw]
&010&&0&\multicolumn{1}{c}{0}&&$\Sv1-\Sv2$\\[\hw]
&&&&&&&$ + \imath S(\Tv1-\Tv2-\Tv3+\Tv4)$\\
\hline
\end{tabular}\\[0.5ex]
with  $C=\cos2\pi lz$ and $S=\sin2\pi lz$
\end{center}}
\end{table*}
proportional to the different combinations of these vectors given in table~\ref{combs}. 
Assuming that the moments are always perpendicular to [001]  an estimate of the absolute magnitude of the magnetic contributions ${\bf M}$ to the nuclear reflections can be obtained from the intensity asymmetry $P$. If ${\bf M}\cdot{\bf M}^*\ll {|F_n|}^2$  then
\[ P=2 (M_y(k_x^2 + k_z^2)-M_xk_xk_y)/P_iF_n\]
where $M_x$ and $M_y$ are the components of the real part of ${\bf M}$ parallel to [100] and [010] respectively, $k_z,k_y$ and $k_z$ are the direction cosines of the scattering vector,  $F_n$ is the nuclear structure factor and $P_i$ the polarising efficiency. The quantities required for the calculation are given in table~\ref{consta} for the reflections which were measured.
\p
\begin{table}[htb]
   \caption{Constants relating the components of the magnetic structure factor to the intensity asymmetry P for several reflections.} \vspace{0.5ex}\begin{center}
 \begin{tabular}{rrcrrrr}  
   		\hline
 $  h\   k\   l$&$k_x^2+ k_z^2$&$k_xk_y$&$F_n$(\mub)&$4f_{Cu}$
   &$8Cf_{Gd} $&$8Sf_{Gd}$\sstrut\\ 
   		\hline
   1\   1\   1& 0.5513& 0.4543& -4.0719& 3.4894& -3.7962&  6.0438\\
   1\   1\   3& 0.7462& 0.2582& 38.9524& 3.1567&  6.5163&  0.7310\\
   2\   1\   0& 0.8033& 0.4008& -3.7710& 2.9723&  6.2291&  0.0000\\
   2\   1\   1& 0.8110& 0.3849&  4.7566& 2.9371& -3.2796&  5.2213\\
   2\   1\   2& 0.8314& 0.3434& -6.0364& 2.8349& -2.5969& -5.3881\\
   2\   1\   3& 0.8572& 0.2912&  4.7566& 2.6753&  5.6540&  0.6343\\
		\hline    
 	\end{tabular}
   \label{consta}
   \end{center}
\end{table}
The 211 reflection is of type A, and comes from the A twin only. For this twin in the WF phase, the moments are nearly parallel and antiparallel to the propagation vector 100 so $\Sv1-\Sv2$ is parallel to $x$. The intensity asymmetry for 211 is positive above 15~K with the magnetic field in the [010] direction. Then, since  $k_xk_y/F_n$ is also positive for 211, $\Sv1-\Sv2$ must be negative to give positive $P$. This shows that for the structure of figure~\ref{orthcell} in which the O$^{2-}$ square coordinating the Cu$^{2+}$ ion at 000 is rotated clockwise, the major component of this Cu$^{2+}$ moment must lie in the negative $x$ direction with the small positive $y$ component parallel to the applied field (see bottom LH panel of figure~\ref{phases}). This shows that the weak ferromagnetism is due to a small rotation of the Cu$^{2+}$ moments in the same direction as that of their coordinating oxygen squares.
\p
The magnetic contributions  ($PF_n/2$)  to the structure factors of the 111 and 113 reflections at 0.5~T and 16~K correspond to -0.53 and 0.75 \mub/cell respectively whereas the magnetisation measured by Seaman et al \cite{seaman:89} amounts to only about $2\times 10^{-3}$ \mub/cell. The intensity asymmetry in these reflections cannot therefore arise from moments aligned parallel to the field direction [010] and must be due to uncompensated moments in the perpendicular direction  [100]. Using the form factors and geometric constants given in table~\ref{consta}, the moments required are 0.47\mub/Gd atom and -0.03 \mub/Cu. The 
magnetic contributions to the  210, 211, 212 and 213 reflections are 0.92,  0.82, 0.55 and 0.32 \mub/cell respectively. They can only be accounted for consistently if the moments $\Sv1-\Sv2$ of both twins, at this temperature and field, lie in the [100] direction and the contribution from gadolinium is small.
These results lead to the model for the magnetic structures of the WF, AFM1 and SF1 phases shown in figure~\ref{phases}. Strong anti-ferromagnetic coupling between the Cu ions in 001 planes combined with anisotropy due to the structural distortion leads to the formation of weakly ferromagnetic  CuO$_2$ layers which at temperatures above $\approx 18$~K, couple along the c-axis with their ferromagnetic moments parallel to one another, to give the weakly ferromagnetic WF phase.
On lowering the temperature below 20~K, exchange interactions due to polarisation of the Gd ions lead to a reversal of the sign of the interlayer coupling so that between 18 and 16~K the long-range interlayer order breaks down and consequently the coherent magnetic scattering drops. At lower temperature anti-ferromagnetic interlayer coupling dominates and gives the
AFM1 structure in which the weakly ferromagnetic layers are
stacked anti-ferromagnetically, so that there is no net ferromagnetic moment. The magnetic propagation vector in the A twin changes from 100 to 010 which is not a reciprocal lattice vector of the A cell; the reverse is true for the B twin. Note that the absence of magnetic scattering
in the \{100\} reflections in zero field at any temperature \cite{chattopadhyay91} shows that the change in interlayer coupling is accompanied by a spin reorientation by 90\degree, so that the moment direction remains parallel to the antiferromagnetic propagation vector.
\begin{table}
\setlength{\hw}{-1.3ex}
\setlength{\vvp}{0.8ex}
\caption{Contributions to the magnetic reflections 101 and 102 from the WF and AFM1 phases of \gdc.}
\begin{center}
\begin{tabular}{{cccccc}}
\hline
Twin&Phase&$\pv$&$\hat{\bf S}$&$101$&$102$\sstrut\\
\hline
&WF&100&100&$p_A(4f_cS)^2q^2$&0\\[\hw]
A\\[\hw]
&AFM1&010&010&0&$p_A(4f_cS)^2$\\
\hline
&WF&010&010&0&$p_B(4f_cS)^2$\sstrut\\[\hw]
B\\[\hw]
&AFM1&100&100&$p_B(4f_cS)^2q^2$&0\\
\hline
\end{tabular}\\[0.5ex]
with  $S=|\Sv1|$ and $q^2=\left(1+(\frac c{a})^2\right)^{-1}$
\end{center}
\label{scatph}
\end{table}%

This model provides an explanation for the anomalous temperature dependence of the SA and SB reflections of figure~\ref{intvst}.  The magnetic contribution to the intensity of the 101 and 102 reflections from each twin in each phase is indicated in table~\ref{scatph}.
If the twin populations are $p_A$ and $p_B=1-p_A$ then the ratio $R_1$ between the mean intensity of 101 in the range 12-15~K to that in t\begin{table}[htdp]
\caption{Twin ratios $\eta$ and magnetisation ratios $R_{\mu}$ calculated from the data of figure~\ref{intvst}} 
\begin{center}
\begin{tabular}{clll}
\hline
Data from fig. \ref{intvst}&\multicolumn{1}{c}{a (d9$_1$)}&\multicolumn{1}{c}{b (d3)}&\multicolumn{1}{c}{c (d9$_2$)}\\
\hline
$R_1$&           $\msp0.73(2)$& $\msp1.06(6)$& $\msp1.74(3)$\\
$R_2$&           $\msp1.15(2)$& $\msp0.980(11)$&$\msp0.88(2)$\\
$\sqrt{R_1/R_2}$&$\msp0.795(13)$&$\msp1.04(3)$&$\msp1.40(2)$\\
$\eta$&         $ -0.114(8)$&$\msp0.020(14)$& $\msp0.168(6)$\\
$R_1R_2-1$&          $-0.17(3)$&$\msp0.04(6)$& $\msp0.54(4)$\\
$R_{\mu}$&       $\msp0.042(7)$&$  -0.010(14)$&$ -0.135(10)$\\
\hline
\end{tabular}
\end{center}
\label{ratios}
\end{table}%
he range 20-30~K should be $\approx p_A/p_B$ and for the 102 reflection $R_2\approx p_B/p_A$. However the products $R_1R_2$ were found to be significantly different from unity which is expected when the Gd sublattices become polarised. If the Gd moments have the magnetic space-group of the AFM1 phase and $U_1$ and $U_2$ are the ratios of the unitary magnetic structure factors for Gd and Cu of the 101 and 102 reflections respectively, then the ratio
$F_{\mu}$ of the Gd to Cu moments can be calculated using 
\[R_1R_2=(1+F_{\mu}U_1)^2(1+F_{\mu}U_2)^2\approx 1+2F_{\mu}(U_1+U_2) \]
 so long as $F_{\mu}<<1$,  in which case  $p_A/p_B\approx\sqrt{R_1/R_2}$. 
Table~\ref{ratios} shows the twin ratio $\eta=(p_A-p_B)/(p_A+p_B)$ obtained using this relationship. The twin ratios $\eta$ and magnetisation ratios $R_{\mu}$ found in the three experiments are significantly different which suggests that the twin populations can change when
the Cu$^{2+}$ ions order. The ratio $R_{\mu}$ can be seen to depend on the twin ratio $\eta$ indicating that the extent to which the Gd moments are polarised in the AFM1 phase somehow depends on the degree of imbalance in the twin populations. 
 \begin{figure}[htbp]
    \centering
  	 \resizebox{0.35\textwidth}{!} {\includegraphics{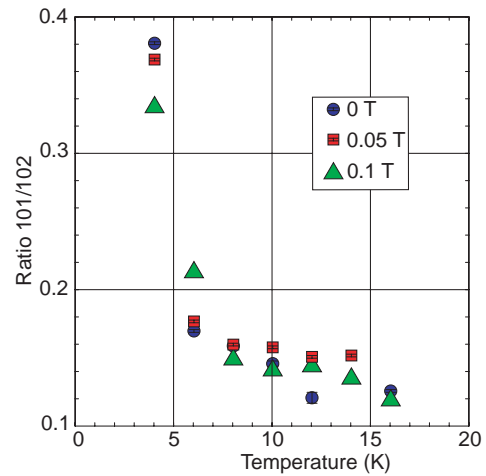}}
    \caption{(b) Temperature variation of the ratio between the 101 and 102 peak heights in  0, 0.05 and 0.1 T.}
    \label{pratios}
 \end{figure}
 
Figure~\ref{pratios} shows the temperature dependence of the ratio between the peak heights of the 101 and 102 reflections measured in  0, 0.05 and 0.1 T.
The absence of any field dependence of this ratio in the WF phase, between 18 and 25~K, shows that the direction of the magnetic moments are coupled sufficiently strongly to the structural
distortion that the weak ferromagnetic moment of the the A twin is not rotated into the [010] direction by a field of 0.5~T. The weak hysteresis observed in the magnetisation measurements \cite{stepanov:93} is then due just to alignment of the 180\degree\ domains in the B twin. 
 At temperatures between 18 and 7~K  the energy gain from reorienting the weak ferromagnetic moments parallel to a magnetic field in the 001 plane may outweigh the inter-layer anti-ferromagnetic coupling energy. In this case a gradual transition with increasing field, or temperature, takes place. With the field applied along [010] the AFM1 structure of the A twin
reverts to the WF structure, but in the B twin the moments in alternate planes rotate by $180$\degree\ to give the SF1 structure. $\Sv1-\Sv2$ for both twins now lies in the [100] direction and the ratio of the 101/102  intensity therefore increases with increasing field as observed. The SF1 structure proposed would give magnetic scattering in the  010 reflection but this cannot be observed in normal beam geometry with [010] parallel
to the $\omega$ axis.
\begin{figure}[htb]
   \centering
 	 \resizebox{0.5\textwidth}{!} {\includegraphics{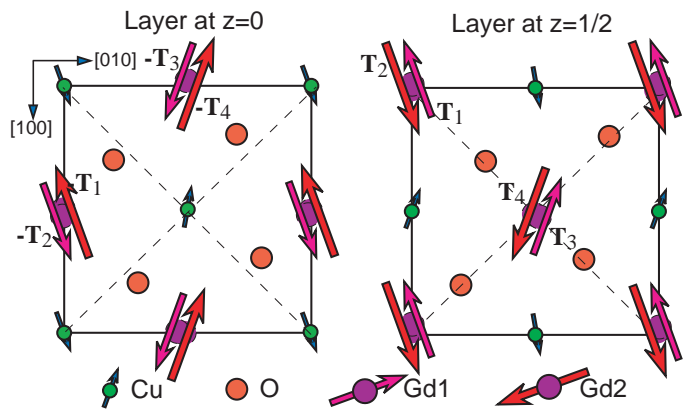}} 
   \caption{Schematic representation of the structure proposed for the   AFM2 phase of \gdc\ in the B twin. The Gd1 atoms are at height $\hf-z_{Gd}$ above the layers and the Gd2 atoms a similar distance below. $z_{Gd}=0.3393$.}
   \label{afm2struc}
\end{figure}
\p
Below 7~K as the Gd sub-lattices order; there is a rapid increase in intensity
in both the fundamental and the 101 and 102 reflections showing that the Gd and Cu sub-lattices have the same propagation vector. Since the magnetic symmetry of the AFM1 structure of figure~\ref{phases},
$Acam'$ for the A twin, is incompatible with the ordering found for Gd in which Gd atoms at $(0,0,\pm z)$ are anti-parallel \cite{chatt:91},  the magnetic space group of the AFM2 phase cannot have a centre of symmetry at the origin.
It can be at most $Aba2'$ in which case the Gd sites at $(0, 0, \pm z_{Gd})$ are no longer equivalent and the moment $|\Tv1|$ need not be equal to $|\Tv2|$. 
The polarisation dependence of the fundamental reflections shows that
in a field ${\bf H} \| [010]$, $\Tv1+\Tv2+\Tv3+\Tv4\ \|\ [100]$. In the AFM2 phase  the magnetic scattering in the F reflections is much greater than that in the A and B types, showing that $\Tv1-\Tv2+\Tv3-\Tv4 \gg \Tv1-\Tv2-\Tv3+\Tv4$ but the behaviour of the 101 and 102 intensity ratio below 7~K shows that the latter sum  $\Tv1-\Tv2-\Tv3+\Tv4\ \|\ [010]$ in the B twin. All these constraints can be satisfied by the structure given in figure~\ref{afm2struc} for the B twin although neither the difference in magnitude of the moments on Gd1 and Gd2 nor their inclination to the crystal axes can be determined precisely from the present data. 

\section{Summary and conclusions}
The various magnetic structures which occur in the H-T phase diagram of \gdc\ are all
built from CuO$_2$ layers in which the Cu$^{2+}$ moments are anti-ferromagnetically coupled. In the ideal NdCu$_2$O$_4$ structure the Cu$^{2+}$ ions are coordinated by squares of O$^{2-}$ ions lying in the planes, but in \gdc\ there is a structural distortion in which the O$^{2-}$  squares around alternate Cu$^{2+}$ sites  rotate in opposite directions. The sense of the polarisation dependence of 
the intensity of Bragg reflections with both magnetic and nuclear contributions shows that the Cu$^{2+}$ moments rotate in the same direction as the oxygen squares. In the temperature range 285-20~K the weakly ferromagnetic  CuO$_2$ layers are stacked with their ferromagnetic moments parallel to one another to give the WF phase. The interlayer coupling is weak and at lower temperature increasing polarisation of the Gd ions which lie between the layers favours antiparallel interlayer coupling and the AFM1 structure which has no
net ferromagnetic moment is stabilised below 18~K. Even a weak magnetic field (0.05~T) can reverse this transition in the favourably oriented twin and induce a spin flop transition to the SF1 phase in the other.  

The unusual temperature dependence of the intensity of the magnetic reflections in \gdc\ which show minima at $T_{C1}\approx 18$ K and at $T_{C2} \approx 8$ K suggests that three-dimensional long-range  magnetic order is temporarily lost at these two temperatures. Two-dimensional magnetic order would be expected to persist and should give rise to corresponding rod-like scattering.  However, experimental measurement of such  diffuse neutron scattering is not possible with the present \gdc\ crystal which has a high  absorption cross section even though partially enriched with $^{158}$Gd.

\end{document}